Superconductivity of bulk MgB$_2$ + *nano*(*n*)-SiC composite system: A high field magnetization study

Arpita Vajpayee[1,2], V. P. S. Awana[1, *], G. L. Bhalla[2] and H. Kishan[1]

[1]Superconductivity and Cryogenics Division, National Physical Laboratory, Dr K.S. Krishnan Road, New Delhi-110012, India

[2]Deparment of Physics and Astrophysics, University of Delhi, New Delhi-110007, India

**Abstract**

We study the effect of *nano*(*n*)-SiC addition on the crystal structure, critical temperature ($T_c$), critical current density ($J_c$) and flux pinning in MgB$_2$ superconductor. X-ray diffraction patterns show that all the samples have MgB$_2$ as the main phase with very small amount of MgO, further with *n*-SiC addition the presence of Mg$_2$Si is also noted and confirmed by SEM & EDS. The $T_c$ value for the pure MgB$_2$ is 18.9K under 8 Tesla applied field, while is 20.8K for the 10-wt % *n*-SiC doped sample under the same field. This points towards the increment in upper-critical field value with *n*-SiC addition. The irreversibility field ($H_{irr}$) for the 5% *n*-SiC added sample reached 11.3, 10 and 5.8 Tesla, compared to 7.5, 6.5, and 4.2 Tesla for the pure MgB$_2$ at 5, 10 and 20K respectively. The critical current density ($J_c$) for the 5-wt % *n*-SiC added sample is increased by a factor of 35 at 10K and 6.5 Tesla field and by a factor 20 at 20K and 4.2 Tesla field. These results are understood on the basis of superconducting condensate (sigma band) disorder and ensuing intrinsic pining due to B site C substitution clubbed with further external pinning due to available *n*-SiC/Mg$_2$Si pins in the composite system.

* Corresponding author: e-mail: awana@mail.nplindia.ernet.in

**Introduction**

MgB$_2$ superconductor has been regarded as a promising material for practical applications at around 20K because of its high critical temperature ($T_c$), large coherence length ($\xi$), simple crystal structure, low material cost, high critical current density ($J_c$) and weak-link-free grain coupling [1,2]. Extensive research has been done on the fabrication of this superconductor in various sample forms, such as polycrystalline bulk, single crystals, metal clad tapes and wires, and thin and thick films [3-5]. For most of the practical applications high critical current density ($J_c$) in the presence of magnetic field along with high upper critical field ($H_{c2}$) and high irreversibility field ($H_{irr}$) are required. $J_c$ of undoped MgB$_2$ is high enough at low magnetic fields for practical application, however, $J_c$ drops rapidly with increasing magnetic field due to low $H_{c2}$ and lack of the effective pinning sites in MgB$_2$. Therefore the improvement of $J_c$ under magnetic field is indispensable for development of MgB$_2$ material for magnet applications. An effective way to improve the flux pinning is to introduce the flux pinning centers into MgB$_2$ through the dopant having size comparable to the coherence length (of the order of few *nm*) of MgB$_2$.

It has already been established by now that moderate impurity doping in MgB$_2$ is effective to increase $J_c$ through introduction of flux pinning centers and/or enhancement of $H_{c2}$ [6-14]. Out of various elements and compounds being doped in MgB$_2$, the carbon-containing compounds such as SiC, C and B$_4$C had been found to be more effective. Here, we are reporting the effect of *nano*-SiC addition on the superconducting properties of MgB$_2$ in high magnetic fields. We found that addition of *n*-SiC into MgB$_2$ helped in significantly enhancing the $J_c$ and $H_{c2}$ in high fields with only slight reduction of $T_c$. This is due to the co substitution of broken *n*-SiC for B in MgB$_2$ lattice inducing intra-grain defects clubbed with high density of nanoinclusions as effective pinning centers. *Nano*-SiC doping is known to enhance both $H_{c2}$ and flux pinning through multiple scattering channel [15]. In the current article, we revisit the MgB$_2$+*nano*(*n*)-SiC composite system, and compare our results with the existing literature. Recently in ref. 16, we reported the low field (< 7 Tesla) magnetization of some of the presently studied MgB$_2$+*nano*(*n*)-SiC samples at 10 and 20K only, and inter-compared the same with $MgB_{2-x}C_x$. This was to highlight the role of C in performance of $MgB_2$ superconductor. In present article we are reporting the detailed high field (up to 14 Tesla) spectacular superconducting performance of though the same system but at various temperatures of 5, 10 and 20K and for a complete series of samples. Further the important superconducting parameter $H_{irr}$ (irreversibility field), which could not be seen in earlier [16] study due to limitation of the applied field, is achieved in the present study. Further it is observed that high field (up to 14 Tesla) performance of *n*-SiC added $MgB_2$ is more spectacular than the low field studies. Our results clearly substantiate the view that the *nano*-SiC addition improves profoundly the high field superconducting performance of $MgB_2$ superconductor.

**Experimental details**

Our polycrystalline $MgB_2$-$nSiC_x$ (x = 0, 3%, 5%, 7% & 10%) samples were synthesized by solid-state reaction route. The Mg powder used is from *Reidel-de-Haen* and amorphous B powder is from *Fluka* (of assay 95-97%). The *n*-SiC powder is from Aldrich having average particle size (*APS*) of 5-12 nm. For synthesis the samples, the stoichiometric amounts of ingredients were ground thoroughly, palletized using hydraulic press and put in a tubular furnace at $850^0C$ temperature under the flow of argon gas at ambient pressure. This temperature was hold for 2.5 hours, and subsequently naturally cooled under the same atmosphere of argon to room temperature. The X-ray diffraction pattern of the compound was recorded by using $CuK_\alpha$ radiation. The scanning electron Microscopy (*SEM*) studies are carried out on these samples using a Leo 440 (Oxford Microscopy: UK) instrument. The magnetoresistivity, $\rho(T)H$, was measured with *H* applied perpendicular to current direction, using four-probe technique. The magnetization measurements were carried out on *Quantum Design PPMS*, equipped with *VSM* attachment.

**Results and Discussions**

The x-ray diffraction (XRD) patterns of $MgB_2$:$n$-$SiC_x$ (x = 0%, 3%, 5%, 7% & 10%) are shown in fig. 1. In case of pure $MgB_2$ all characteristic peaks are obtained and their respective indexing is shown in the figure itself. The structure of $MgB_2$ belongs to space group *P6/mmm*. It can be seen that all doped samples along with the undoped exhibit well developed $MgB_2$ phase, with only a small amount of MgO present, which is marked by symbol 'o' in fig. 1. The presence of small quantity of MgO being present with $MgB_2$ main phase is consistent with earlier reports on similar samples [17-19]. No other impurity phases like $Mg_2C_3$ and $MgB_2C_2$ are detected.

As we go on increasing the added $n$-SiC content in $MgB_2$:$n$-$SiC_x$, the presence of $Mg_2Si$ (*) and unreacted SiC (+) is noticed. At lower doping level (x < 7wt%) the sample consist of a major phase of $MgB_2$ with minority phase of $Mg_2Si$ and as we increase (x ≥ 7wt%) the doping level of $n$-SiC, the amount of this non-superconducting phase was increased. In particular for $MgB_2$:$n$-$SiC_x$ (x = 7% & 10%), the {hkl} planes {111}, {220} and {400} of the $Mg_2Si$ are noticed clearly in Fig.1. Presence of $Mg_2Si$ has earlier been seen in a recent report [20]. As far as the majority $MgB_2$ phase is concerned, the peak situated between $2\theta = 33^o$ and $2\theta = 34^o$ shifts towards the higher $2\theta$ values with increasing x, indicating the contraction in a-axis in crystal lattice. The lattice parameters, 'a' and 'c', of the hexagonal $AlB_2$ type structure of $MgB_2$ are calculated using these peak shifts, and their variation is tabulated in Ref. [16]. The decrease in 'c' parameter with increasing x (content of $n$-SiC) is relatively small as compared to 'a' parameter. The variation in 'a' parameter indicated the partial substitution of B by C [21,22].

The grains morphology of pure and 10wt%-$n$SiC added $MgB_2$ is shown in Fig. 2. The pristine $MgB_2$ grains are of average size with in less than a micron, also seen are some porous/MgO insulating white regions, see Fig. 2(a). The presence of $Mg_2Si$ can be noticed in 10wt%-$n$SiC added $MgB_2$ as large spherical white regions in the $MgB_2$ matrix along with smaller insulating MgO. The elemental analysis of these micrographs showed the presence of Si in 2(b) but not in 2(a). This is consistent with the fact that unreacted free Si is present in $n$-SiC doped samples to form the desired $Mg_2Si$ phase, being seen in XRD in Fig. 1. Though, all the elements being present in $MgB_2$ or $n$-SiC doped samples such as Mg, B, and O in former and in addition Si, and C in later are seen in EDS, the actual percentage ratio is not determined because of very light element Boron and Carbon in comparison to others. The sensitivity of SEM is known to be relatively poor for lighter elements such as B, C and O.

Fig. 3 shows the resistance versus temperature curves under magnetic field $R(T)H$ up to 8 Tesla for the undoped, 5% $n$-SiC and 10% $n$-SiC doped samples. The transition temperature ($T_c$) for the pure sample is 38.1K in zero applied field. For the 10 wt% $n$-SiC added sample $T_c$ decreased to 34.5K in zero applied field. Further, it is noted that the $R(T)$ curves for the doped samples shifted with increasing magnetic field much more slowly than the pure one. The $T_c$ value for the pure $MgB_2$ is 18.9K for 8 Tesla applied field while is 20.8K for the 10-wt % $n$-SiC doped sample under the same field.

A further important point is that the nominal resistance of these samples is very different, $R(40K)$ being 290 µΩ for the undoped sample, 490 µΩ for 5 wt % and 800 µΩ for the 10 wt % $n$-SiC doped sample. This is shown in fig. 4. It refers that the scattering increases with increasing $n$-SiC content. In inset of fig. 4 the temperature dependence of normalized resistance $R(T)/R(275K)$ is shown for pure and 10% doped sample. The residual resistivity ratio (RRR = $R_{T275K}/R_{Tonset}$) values for the pure and 10% SiC doped samples are 3.15 and 1.74 respectively. Both C doping (revealed by contraction in 'a' parameter and reduction in $T_c$) and the inclusion of $Mg_2Si$ (revealed by XRD) can enhance the electron scattering, and hence the decreased RRR values. Further, the higher values of room temperature resistivity for doped samples indicate that the impurity scattering is stronger due to the Carbon substitution at Boron sites. This is in agreement with previous studies on $MgB_{2-x}C_x$ systems [21,23]

The variation of upper critical field with respect to reduced temperature is shown in fig. 5, with the help of resistive transitions shown in Fig. 3. All the doped samples show the higher values of critical field in comparison to the pure sample at all the temperatures. The *n*-SiC reacts with the Mg and releases highly reactive free C on the atomic scale at the same temperature where formation of $MgB_2$ takes place. Because of the availability of reactive C atom at that time, the C can be easily incorporated into the lattice of $MgB_2$ and substitute into B sites [24]. The carbon substitution into boron site in lattice is responsible for creating the disorder on the lattice site of boron, which leads to the enhancement in value of upper critical field. Further it is worth mentioning that magnetic field, which is required to destroy the *bulk* superconductivity, is smaller than the magnetic field needed to destroy the *surface* superconductivity, the later is 1.6946 ($\eta$) times higher than the former [25,26]. Although this ratio ($\eta$) varies with temperature for two- band superconductors but in ref. 25 *Denis* discussed that at temperature near about the transition temperature ($T_c$) $\eta$ takes its highest value that is near about 1.7; overall $\eta$ takes value from $\approx$1.7 to $\approx$1.64 for temperature variation from $T_c$ to down towards 0K. Therefore, in fig. 5 we are showing the variation of upper critical field $H_c$, where $H_c = 1.7 \times H$ at $R \rightarrow 0$ ($T \rightarrow T_c$), against the reduced temperature.

The magnetic hysteresis loop for all the doped samples $MgB_2$:*n*-$SiC_x$ (x = 0%, 3%, 5%, 7% & 10%) are shown in fig. 6 at *T* = 5, 10, 20K and under up to 13 Tesla applied field. This figure clearly demonstrates that at *T* = 5K the closing of *M(H)* loop for pure sample is at 7.5 Tesla, while the same is closed at 11.3 Tesla for 5% *n*-SiC doped sample. This indicates that there is quite an improvement in irreversibility field values by addition of *n*-SiC in parent compound. The irreversibility fields ($H_{irr}$) are derived from the fields at which the magnetic hysteresis loop gets nearly closed; with the criterion of giving the $J_c$= 100 A/cm$^2$. To know the effect of doping level of *n*-SiC on $H_{irr}$ values a plot is drawn in $H_{irr}$ versus *x* (concentration of *n*-SiC) and it is shown in fig. 7 at 5K, 10K & 20K. Doping with *n*-SiC has significantly improved the $H_{irr}$. At all the temperatures the 5% *n*-SiC doped samples gives the best value of $H_{irr}$. The values of $H_{irr}$ for the 5% *n*-SiC added sample reached 11.3, 10 & 5.8 Tesla, compared to 7.5, 6.5, 4.2 Tesla for the pure one at 5, 10 &

20K respectively. Worth mentioning is the fact that in an earlier preliminary study [16] by some of us, the closing of $M(H)$ loops for $n$-SiC added samples could not be achieved due to non availability of higher applied fields. The spectacular enhancement in $H_{irr}$ values, being seen in Fig.7, is definitely due to improvement in flux pinning in $MgB_2$ by the $n$-SiC doping.

The magnetic $J_c$ for all the samples was calculated from the $M(H)$ loop at 5, 10 & 20K. Fig. 8(a) & 8(b) show the magnetic $J_c$ vs $H$ for all the samples at 20 & 10K respectively. At low fields all the samples attain about $10^5$ A/cm$^2$ $J_c$ at both the temperatures. Among all the doped samples 5% $n$-SiC doped sample gives the best performance. For undoped sample $J_c$ drops rapidly in the presence of magnetic field and is almost negligible above 4.2 Tesla and at 6.5 Tesla at 20K and 10K respectively. On the other hand the 5% $n$-SiC doped sample exhibits the $J_c$ of the order of $10^3$A/cm$^2$ at the corresponding fields at both the temperatures. The $J_c$ is 35 times higher than pure one at 10K in 6.5 Tesla field in case of 5% $n$-SiC doped sample and 20 times higher at 20K in 4.2 Tesla for the same sample. Because of the dual reaction [24], first reaction of $n$-SiC with Mg forming $Mg_2Si$ and second free C being incorporated into $MgB_2$ both helps in pinning of vortices and improved superconducting performance. $Mg_2Si$ and excess carbon can be embedded within the grain of $MgB_2$ as nanoinclusions. Due to the substitution of C at B site the formation of nanodomain structure takes place due to the variation of Mg-B spacing. These nanodomains defects having the size of 2-3 nm can also behave as effective pinning centers. So, the highly dispersed nanoinclusions within the grains and the presence of nanodomain defects are acting as pinning centers and thus resulting in the improved $J_c(H)$ behavior for the $n$-SiC doped samples.

To confirm the improved flux pinning behavior through SiC doping, the field dependence of normalized flux pinning force ($F_p / F_{p, max}$) is shown in fig. 9(a) & 9(b) at 20K & 10K. The relationship between flux pinning force and critical current density could be described by [27,28]

$$F_p = \mu_0 J_c(H) H \qquad (1)$$

Where $\mu_0$ is the magnetic permeability in vacuum. These figures depict a significant improvement in pinning forces by *n*-SiC doping at both the temperatures for fields greater than 2 Tesla. Flux pinning curves for the doped samples are shifted to the right as compared to pure $MgB_2$. As described earlier the nanoinclusions and nanodomains having the size comparable to coherence length of $MgB_2$, can work as point pinning centers, causing a shift of the curve in $F_p / F_{p,max}$ vs $H$ curve towards the higher field. It can be seen from fig. 9 that for the 5-wt% *n*-SiC doped sample the peak is much broader than those of other samples, indicating that the highest pinning strength for this sample; this is in confirmation with $J_c$ results. As far as the type of pinning i.e., grain boundary, point defects or order parameter change is concerned; in our case it is seemingly the combination of grain boundary and point defects. The former is due to the presence of grain boundary precipitates of $Mg_2Si$ and later due to inclusions of *n*-SiC. This we presume because the $F_p / F_{p,max}$ vs $H$ peak does not only get broadened (grain boundary pinning) with *n*-SiC addition but there is a slight right direction shift in peak position to higher fields as well [29].

Worth mentioning is the fact that though all *n*-SiC added samples till 10wt% addition are superior than that to the pristine $MgB_2$, the 5-wt% *n*-SiC added one is the optimum. In our recent another paper [16] in which we discussed about role of carbon in $MgB_2$ lattice to enhance the flux pinning performance comparatively at lower fields, we found that 7-wt% *n*-SiC added sample of different batch showed the highest improvement in the flux pinning. Therefore, we can say that due to *n*-SiC doping superconducting performance enhances profoundly and the optimum is found between say 3 to 7-wt% additions. The fact is that 5-wt% or 7-wt% amount of *n*-SiC is not the optimum for achieving the best $J_c$, $H_{irr}$ & $H_{c2}$; it all depend on the various other factors like synthesis temperature, heating rate, annealing time, magnetic field and resultant sample quality etc [7,30].

**Conclusions**

In summary, the effect of *nano* SiC doping on critical temperature ($T_c$), critical current density ($J_c$) and flux pinning was investigated under a wide range of magnetic field. We found that a significant flux pinning enhancement in $MgB_2$ can be easily achieved using *nano* SiC as an additive. The Si and C released from the decomposition of *nano* SiC at the time of formation of $MgB_2$ formed $Mg_2Si$ and substituted at B sites respectively. The C substitution for B resulted in a large number of intra-granular dislocations and dispersed nanosize impurities, which are both responsible for the significant enhancement in flux pinning.

**Acknowledgement**

The authors from *NPL* would like to thank Dr. Vikram Kumar (*DNPL*) for his great interest in present work. Mr. A.K. Sood from *SEM* Division of *NPL* is acknowledged for providing us with the *SEM* micrographs. Dr. Rajeev Rawat from *CSR*-Indore is acknowledged for the resistivity under magnetic field measurements. Mr. Kranti Kumar and Dr. A. Banerjee are acknowledged for the high field magnetization measurements. Further *DST*, Government of India is acknowledged for funding the 14 Tesla-PPMS-VSM at CSR, Indore. Arpita Vajpayee would like to thank the *CSIR* for the award of Junior Research Fellowship to pursue their *Ph. D* degree.

**Figure captions**

Figure 1. X-ray diffraction pattern of pure and *n*-SiC doped samples

Figure 2 (a,b). SEM pictures of Pure $MgB_2$ & 10-wt% *n*-SiC added sample

Figure 2(c). EDS pattern of pure $MgB_2$ and 10-wt% *n*-SiC added sample. Peaks for different elements are marked.

Figure 3. Superconducting transition zone of Resistance vs Temperature plot under applied magnetic field *R(T)H* up to 8 Tesla for pure, 5-wt% & 10-wt% *n*-SiC added samples

Figure 4. Variation of resistance with temperature *R(T)* plots for Pure, 5% and 10% *n*-SiC added samples

Figure 5. Upper critical field ($H_{c2}$) vs normalized temperature plots for $MgB_2$+*n*-$SiC_x$ x=0%, 3%, 5%, 7% & 10% samples

Figure 6. Magnetization loop *M(H)* for $MgB_2$+*n*-SiCx (x=0%, 3%, 5%, 7% & 10%) up to 13 Tesla field at 5, 10 & 20K

Figure 7. Variation of irreversibility field $H_{irr}$ with respect to *n*-SiC concentration at 5K, 10K & 20K

Figure 8. $J_c(H)$ plots for $MgB_2$+*n*-$SiC_x$ samples along with pristine $MgB_2$ at (a) 20K & (b) 10K

Figure 9. Variation of reduced flux pinning force ($F_p/F_{p,max}$) with magnetic field for $MgB_2$+*n*$SiC_x$ (x=0%, 3%, 5%, 7% & 10%) at (a) 20K & (b) 10K

Figure 1

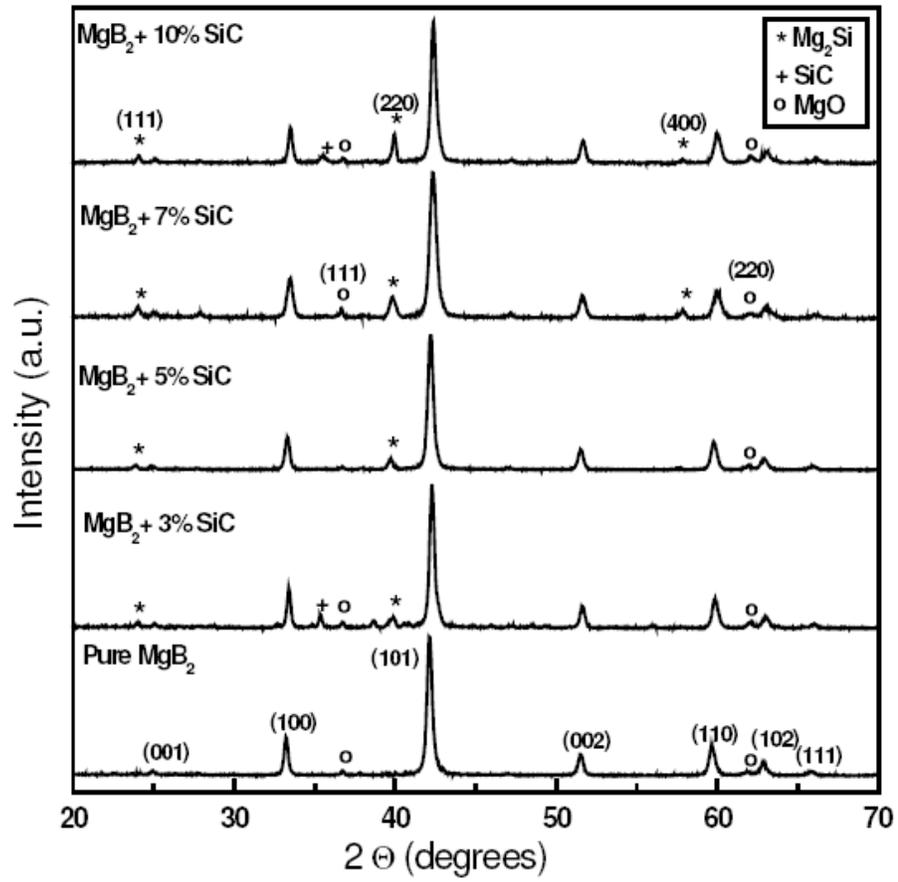

Figure 2(a)

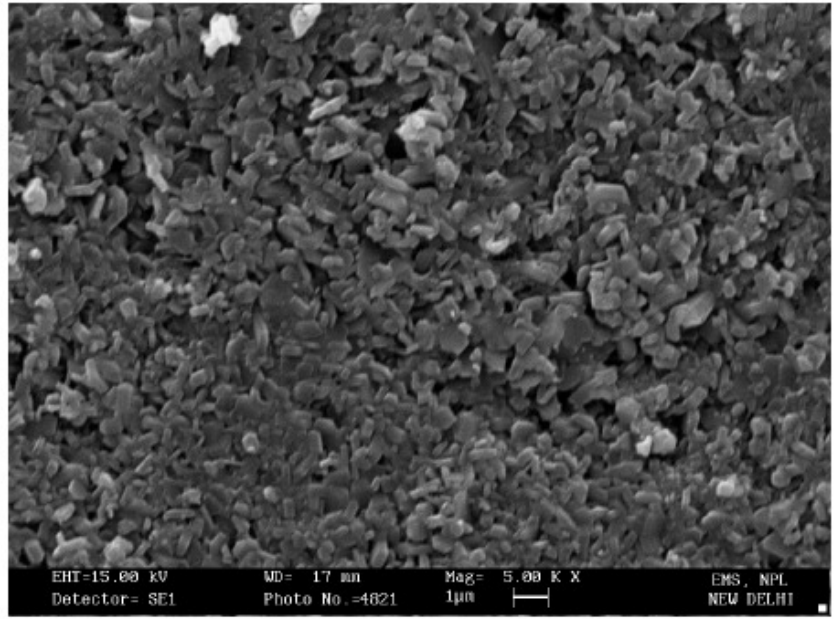

Figure 2(b)

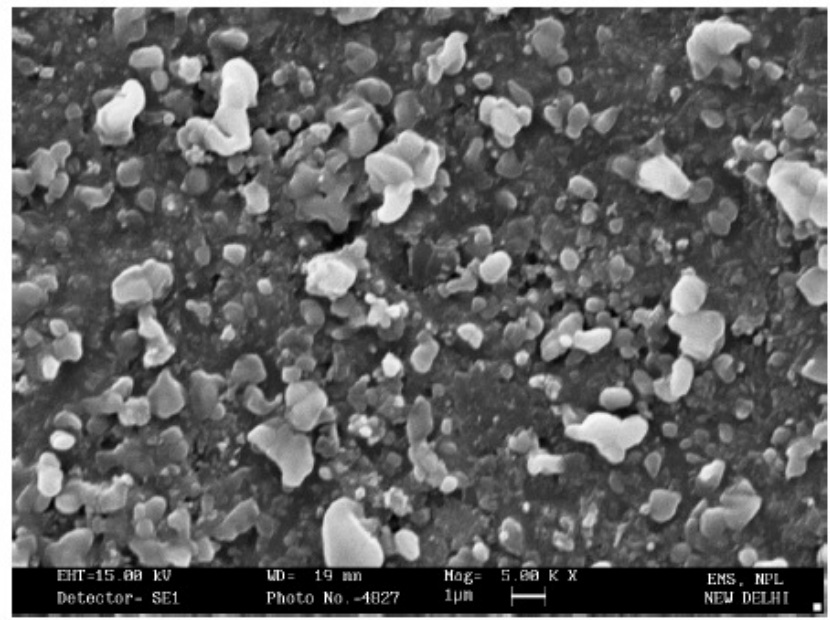

Figure 2(c)

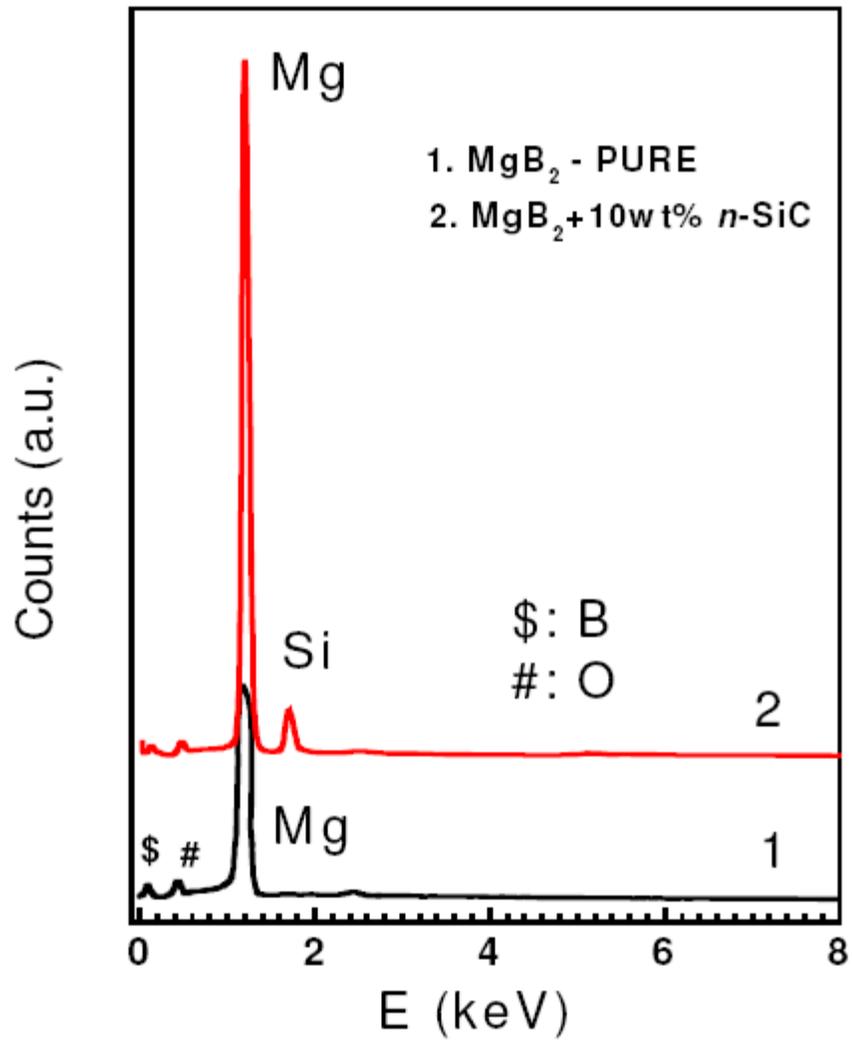

Figure 3

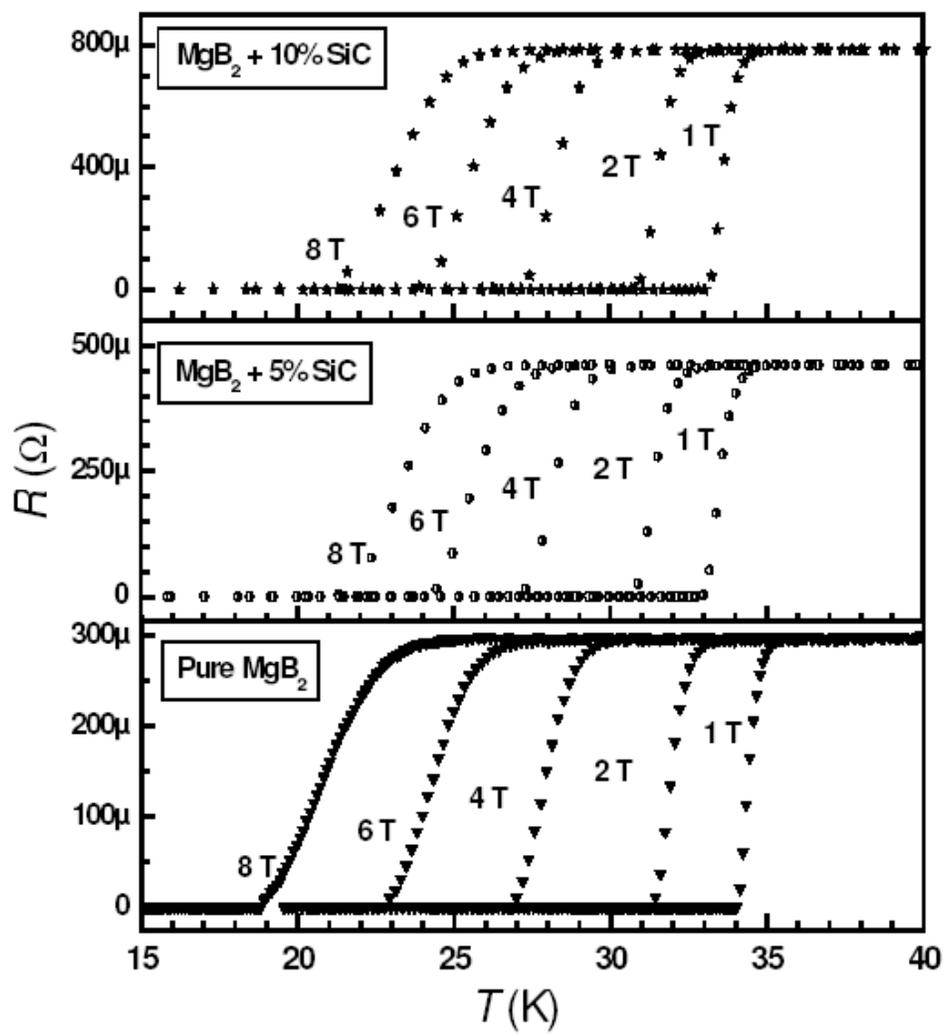

Figure 4

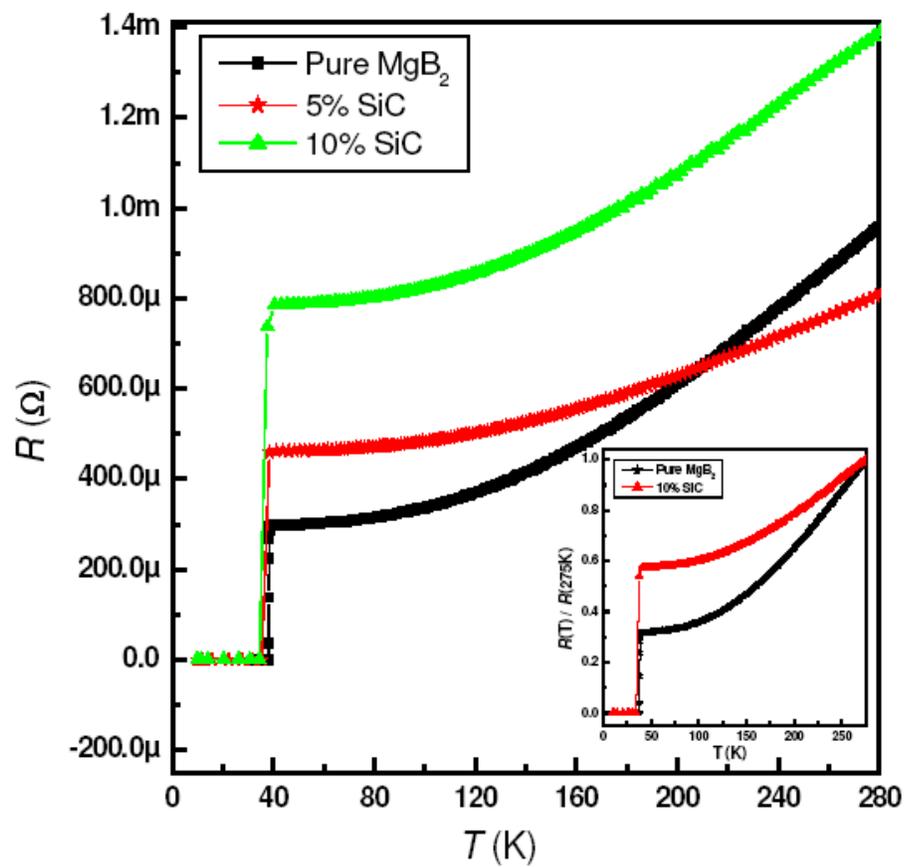

Figure 5

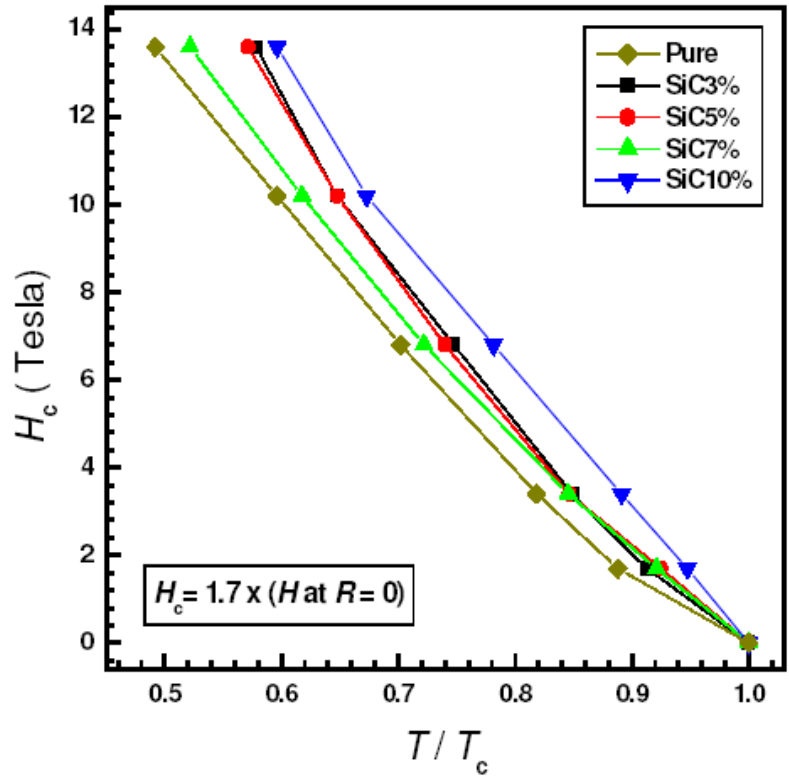

Figure 6

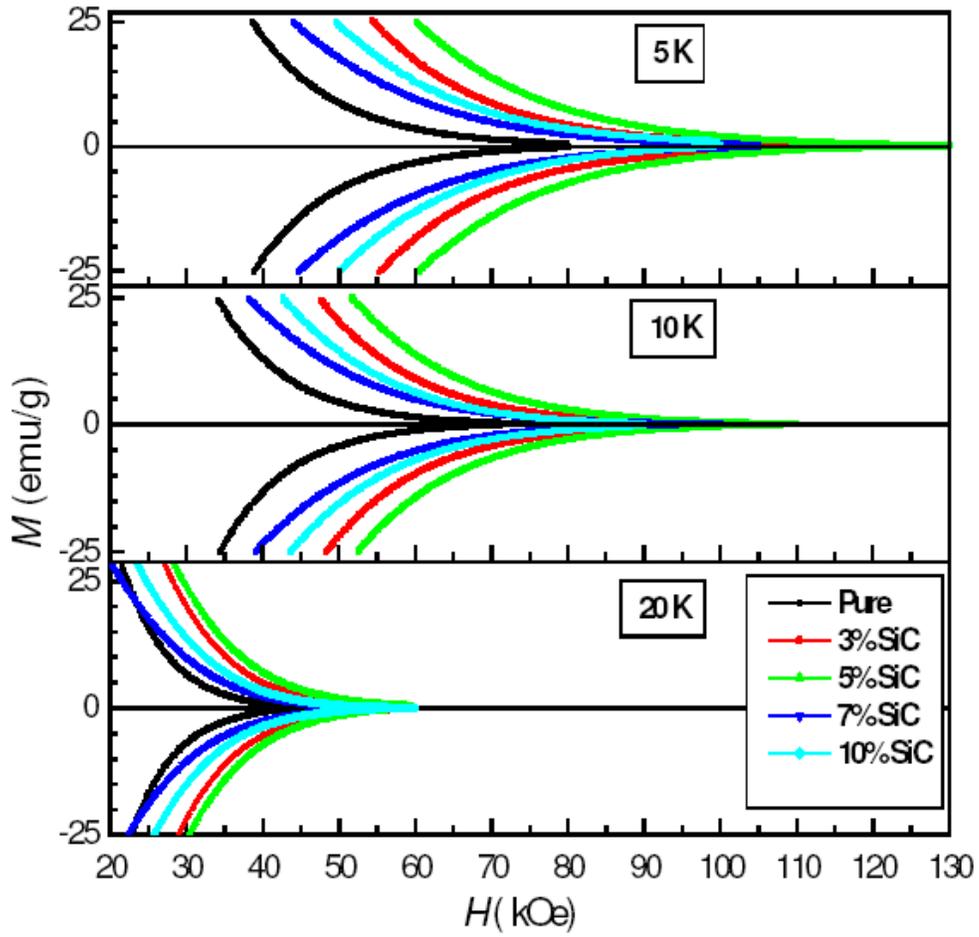

Figure 7

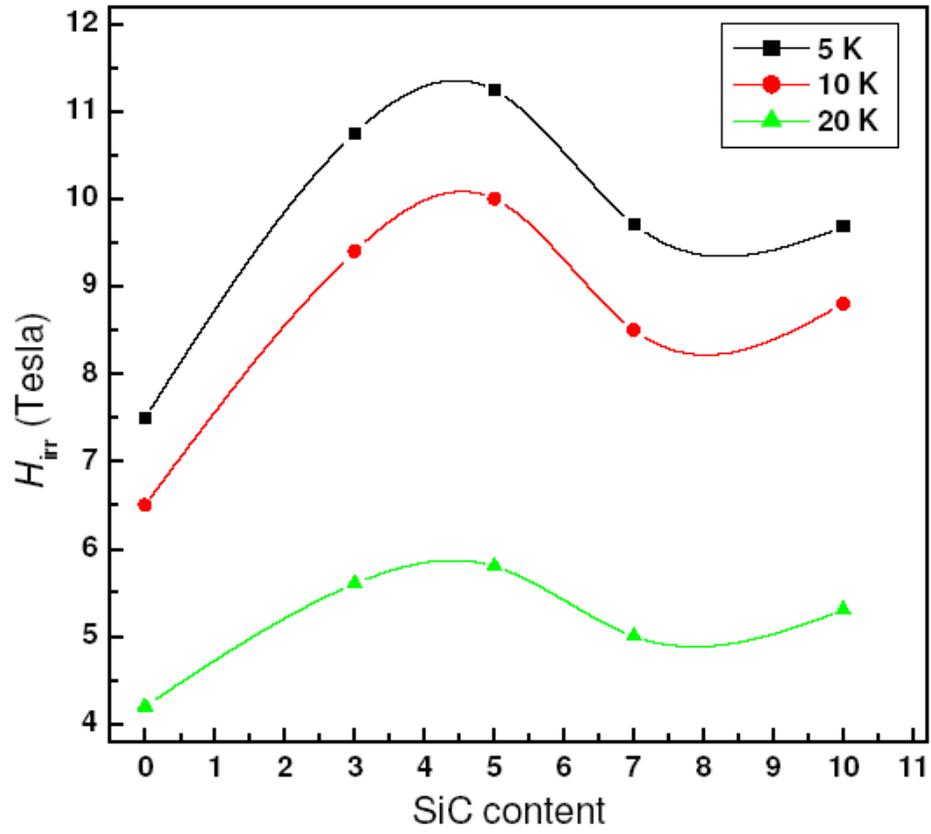

Figure 8 (a) & 8 (b)

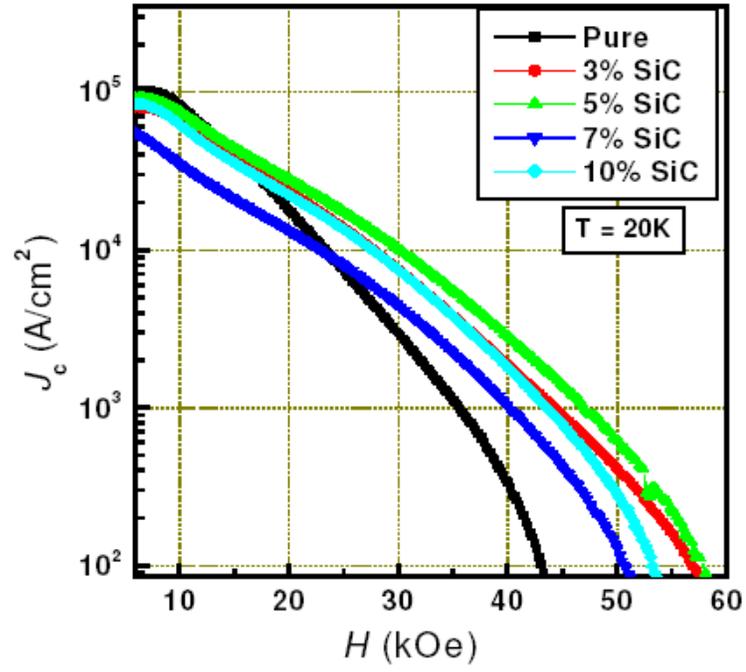

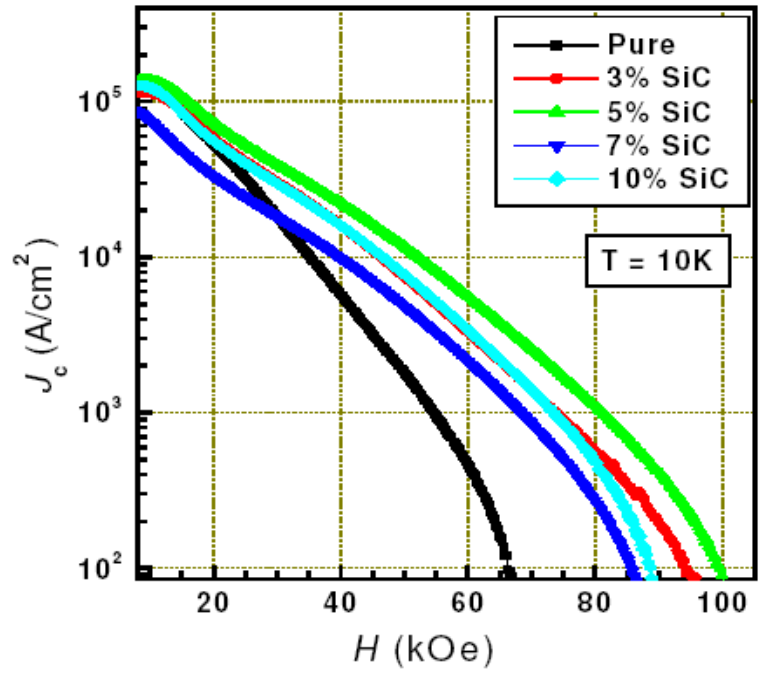

Figure 9(a) & 9(b)

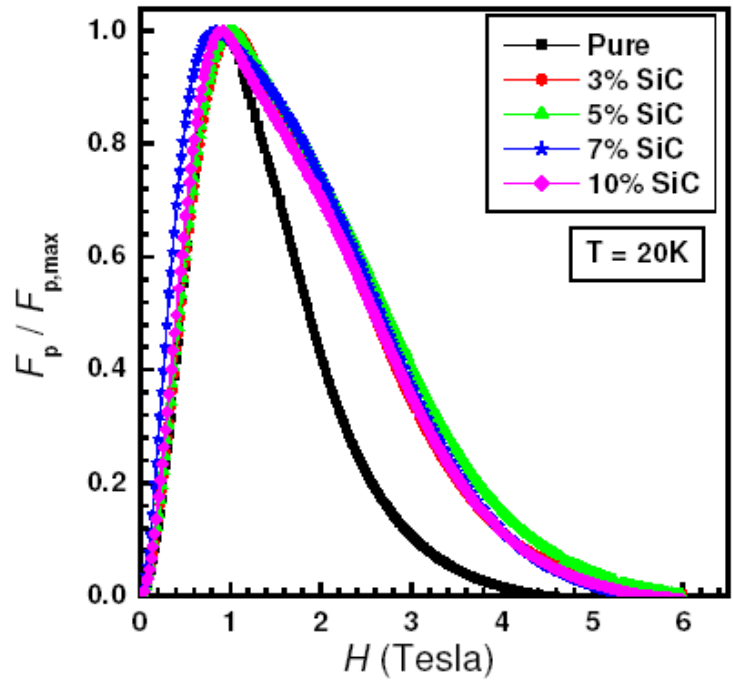

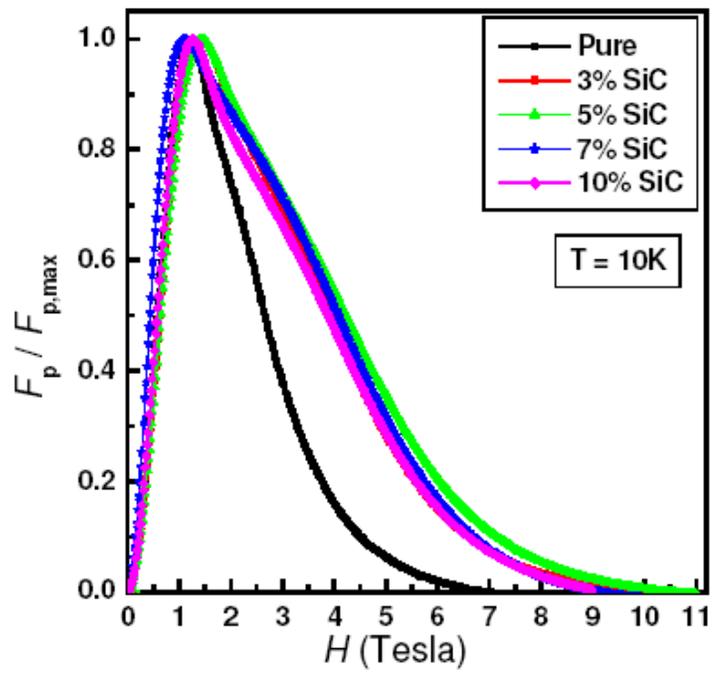